\DeclareSymbolFont{AMSb}{U}{msb}{m}{n}
\documentclass[reqno,centertags,11pt,a4paper,noamsfonts]{amsart}
\pdfoutput=1
\usepackage[svgnames]{xcolor}
\usepackage{graphicx}
\usepackage[english]{babel}

\usepackage[utf8]{inputenc}
\usepackage[babel=true,expansion=alltext,protrusion=alltext-nott,final]{microtype}
\usepackage[margin={3cm},marginparwidth={2.5cm}]{geometry}
\usepackage[
  backend=biber,
  style=numeric,
  sorting=none,
  giveninits=true,
  doi=true,
  maxbibnames=6,
  isbn=false,
  url=true,
  eprint=false,
  sorting=nyt
]{biblatex}

\renewbibmacro{in:}{%
  \ifentrytype{article}
    {}
    {\printtext{\bibstring{in}\intitlepunct}}}

\AtEveryBibitem{%
  \clearfield{note}%
}

\renewbibmacro*{doi+eprint+url}{%
  \printfield{doi}%
  \newunit\newblock
  \iffieldundef{doi}
    {\usebibmacro{url+urldate}}
    {}%
}

\usepackage{textcomp}
\usepackage[sb]{libertine}
\usepackage[libertine,bigdelims,vvarbb]{newtxmath}
\useosf
\usepackage[varqu,varl]{inconsolata}
\usepackage[supsfam=LibertinusSerif-Sup,%
       supscaled=1.2,%
       raised=-.13em]{superiors}
\usepackage{mathtools}
\usepackage[T1]{fontenc}
\usepackage{textcomp}
\usepackage{amsthm}
\usepackage{csquotes}
\usepackage[inline]{enumitem}
\numberwithin{equation}{section}
\usepackage[normalem]{ulem}

\usepackage[cal=cm]{mathalpha}
\usepackage{float}  
\usepackage{caption}
\usepackage{subcaption} 
\usepackage{xspace}         
\usepackage[scientific-notation=true]{siunitx}      
\usepackage{pstricks}
\usepackage{tikz}
\usetikzlibrary{positioning} 
\usetikzlibrary{calc}
\usepackage{pgfplots}
\pgfplotsset{width=10cm,compat=1.9}
\usepackage{bm}
\usepackage{sidecap}
\usepackage{graphicx,color}

\usepackage{amsmath}
\DeclareFontFamily{U}{mathx}{}
\DeclareFontShape{U}{mathx}{m}{n}{<-> mathx10}{}
\DeclareSymbolFont{mathx}{U}{mathx}{m}{n}
\DeclareMathAccent{\widehat}{0}{mathx}{"70}
\DeclareMathAccent{\widecheck}{0}{mathx}{"71}

\usepackage[pagebackref=false]{hyperref}

\newcommand{\R}{\mathbb{R}}
\newcommand{\C}{\mathbb{C}}
\newcommand{\Z}{\mathbb{Z}}
\newcommand{\N}{\mathbb{N}}

\newcommand{\mL}{\mathrm{L}}
\newcommand{\ee}{\mathrm{e}}
\newcommand{\mH}{\mathrm{H}}

\newcommand{\norm}[1]{\lVert #1\rVert}

\newcommand{\inner}[1]{\left\langle #1 \right\rangle}

\newtheorem{theorem}{Theorem}[section]
\newtheorem{proposition}[theorem]{Proposition}
\newtheorem{lemma}[theorem]{Lemma}
\newtheorem{corollary}[theorem]{Corollary}

\theoremstyle{definition}

\theoremstyle{remark}
\newtheorem{remark}[theorem]{Remark}
\newtheorem*{remark*}{Remark}

\hypersetup{
  linktoc=all,
  bookmarksdepth=3,
  colorlinks=true,
  linkcolor=green!40!black,
  citecolor=orange!70!black,
  urlcolor=blue!60!black,
  pdftitle={Ground-state energy is convex in the number of electrons in 1D},
  pdfauthor={Thiago Carvalho Corso}}

\begin{document}
\numberwithin{equation}{section}
\numberwithin{table}{section}

\title[Ground-state energy is always convex in the number of electrons for one dimensional systems]{Ground-state energy is always convex in the number of electrons for one dimensional systems}

\author[T.~Carvalho~Corso]{Thiago Carvalho Corso}
\address[T.~Carvalho Corso]{Institute of Applied Analysis and Numerical Simulation, University of Stuttgart, Pfaffenwaldring 57, 70569 Stuttgart, Germany}

\date{\today}

\begin{abstract}
We prove that for one-dimensional fermions in the real line with a convex and non-increasing interaction potential, the ground-state energy is convex in the number of particles for arbitrary external potentials. As a consequence, we show that any external potential capable of binding $n$ fermions also binds $n-1$ fermions on the line. This settles a well-known conjecture in quantum chemistry in the simplified setting of one-dimensional systems. 
\end{abstract}

\maketitle
\setcounter{tocdepth}{1}
\setcounter{secnumdepth}{2}

\section{Introduction}

\subsection*{Motivation}

A well-known question in quantum chemistry is the following: given a molecule that can bind $n$ electrons, can it also bind $n-1$ electrons? Although the intuitive answer is obviously yes, rigorously deriving this fact from the many-body Schr\"odinger equation remains a major open problem in mathematical physics \cite{Lieb1983, Simon1984Problems, LiebSeiringer2010, DiMarinoLewinNenna2024, Lewin2025}. In fact, experimental atomic ionization data \cite{NIST_ASD_2024} suggest the stronger statement that the ground-state energy is convex in the number of particles, i.e., 
\begin{align} E_{n-1}(v) + E_{n+1}(v) \geq 2 E_n(v), \label{eq:ground-state convexity inequality} \end{align}
where $E_n(v)$ is the ground-state energy of the $n$-electron system under an external electric potential $v$. For atoms, this inequality means that the energy necessary to remove an electron bound to a nuclei should increase as the number of remaining bound electrons decrease. 

This question is related to the ionization problem \cite{Nam2013Ionization,Nam2022Ionization}. Indeed, the ionization problem asks what is the maximal number of electrons a given molecule or atom can bind. Hence, convexity of the ground-state energy (together with the HVZ theorem) implies that this number is the largest integer $n$ such that the ionization energy $E_{n-1}(v) - E_n(v)$ is strictly positive. Moreover, the convexity inequality~\eqref{eq:ground-state convexity inequality} also has interesting consequences in Density Functional Theory (DFT) \cite{PPLB1982,Lieb1983,PerdewLevy1983,ShamSchluter1983,ParrYang1989}. More precisely, this inequality for arbitrary external potentials, is equivalent to the statement that a minimizer of the grand-canonical Lieb functional for a given density with possibly fractional particle number $n=\int \rho \in \R$ can be chosen as a convex combination of density matrix with $\lceil n \rceil$ and $\lfloor n \rfloor$ particles \cite{LewinLiebSeiringer2023,DiMarinoLewinNenna2025}. This observation is particularly interesting in applications as it allows one to compute the ionization energies by studying the regularity properties of the Lieb functional \cite{BaerendsGritsenkoVanMeer2013,MoriSanchezCohen2014,KraislerKronik2014}.

Recently, inequality~\eqref{eq:ground-state convexity inequality} has been established for one-dimensional systems in the \emph{classical} case \cite{DiMarinoLewinNenna2025} , i.e., with zero kinetic energy and for probability measures instead of wavefunctions. This result was obtained by using optimal transport methods, inspired by the so-called strictly correlated (or strongly interacting) limit of DFT \cite{Sei99,CDD15}. On the other hand, in the opposite limit of non-interacting systems, inequality~\eqref{eq:ground-state convexity inequality} is easily seen to be true in any dimension and for any external potential, as the gap $E_n(v) - E_{n-1}(v)$ is simply given by the $n^{th}$ lowest eigenvalue of the single particle operator $-\Delta + v$. However, in the arguably most relevant intermediate regime of interacting electronic systems, this conjecture is not true in general. More precisely, in his celebrated DFT paper \cite{Lieb1983}, Lieb provided a counterexample to this conjecture in the case of hard-core interaction potential. This example was later generalized to certain Riesz interaction potentials by Ayers \cite{Ayers2024}. The first counterexample with Coulomb interaction potentials and Coulomb external potentials with small fractional charges were obtained in~\cite{DiMarinoLewinNenna2024}. Further negative results with Coulomb potentials \cite{MartinezGonzalezAyers2026}, harmonic interactions \cite{LevyLeblond1968, PhillipsDavidson1983}, in the 1D Hubbard model \cite{FyeMartinsScalettar1990} and in nuclear physics \cite{BlaizotRipka1985} have also been found. Nevertheless, to the best of our knowledge, no positive answer to this question for a general class of external and interaction potentials is known so far. 

It is therefore the main goal of this paper to present such a positive answer for one-dimensional spinless fermions.


\subsection{Main result} 

More precisely, the main result of this paper can be stated as follows. For measurable $w:(0,\infty) \rightarrow [0,\infty)$ and $v\in \mH^{-1}_{\rm unif}(\R;\R)$ we define the ground-state energy $E_n(v,w)$ as
\[ E_n(v,w) \coloneqq \inf \left\{ \frac{\inner{\Psi, H_n(v,w) \Psi}}{\norm{\Psi}_{\mL^2(\R^n)}^2}  : \Psi \in \mH^1(\R^n) \cap \wedge^n \mL^2(\R) \right\}, \]
where
\begin{align}
    \inner{\Psi, H_n(v,w) \Psi} \coloneqq \norm{\nabla \Psi}_{\mL^2}^2 + \int_{\R^n} \sum_{j<k} w(|x_j-x_k|) |\Psi(x_1,...,x_n)|^2 \mathrm{d} x_1...\mathrm{d} x_n + \inner{v, \rho_{\Psi}},\label{eq:form definition}
\end{align}
and 
\[ \rho_\Psi(x) = n \int_{\R^{n-1}} |\Psi(x,x_2,...,x_n)|^2 \mathrm{d} x_2... \mathrm{d} x_n\]
is the single-particle density of $\Psi$. Note that, since $v$ is Laplace form-bounded (see Lemma~\ref{lem:potentials} later) and $w\geq 0$, the form~\eqref{eq:form definition} is well-defined (though possibly $+\infty$) for any $\Psi \in \mH^1(\R^n)$. We then have
\begin{theorem}[Convexity of ground-state energy in the particle number] \label{thm:GS convexity} Let $w:(0,\infty) \rightarrow [0,\infty)$ be a convex and non-increasing interaction potential. Then, for any potential $v\in \mH^{-1}_{\rm unif}(\R;\R)$, the following inequality holds:
\[ E_{n+1}(v,w) + E_{n-1}(v,w) \geq 2E_n(v,w) , \qquad \text{for any $n\in \N$,} \]
with the convention $E_0(v,w) = 0$.
\end{theorem}

\begin{remark}[External potentials] The space $\mH^{-1}_{\rm unif}(\R)$ denotes the set of distributions $v\in \mathcal{D}'(\R)$ such that the restriction to each unit interval $I\subset \R$ belongs to $\mH^{-1}(I)$ and
\[ \norm{v}_{\mH^{-1}_{\rm unif}(\R)} = \sup_{t\in \R} \norm{v}_{\mH^{-1}((t,t+1))} < \infty. \]
This is the maximal class of distributional potentials in $\R$ which are Laplace form-bounded \cite{MazyaVerbitsky2005}. 
\end{remark}

\begin{remark}[No singularity constraint on the interaction $w$]
We emphasize that no additional restrictions on the behaviour of $w(r)$ as $r\downarrow 0$ are necessary. In particular, the result still holds for $w(r) = r^{-\alpha}$ for any $\alpha>0$.
\end{remark}

\begin{remark}[Spin vs. spinless fermions for singular interactions]
For non locally integrable interactions such as the Coulomb potential $w(r) = 1/r$, the spin and spinless ground-state energies are the same, as any wave-function with finite  interaction energy has zero trace along the collision set $\{x\in \R^n: x_j = x_k \text{ for some $j\neq k$} \}$ and can therefore be turned into an spatially anti-symmetric wave-function with the same energy, see \cite{Girardeau1960,Cor26b}. In particular, the result from Theorem~\ref{thm:GS convexity} automatically extends to spin systems for such interactions.
\end{remark}

As an immediate corollary of this result, we obtain the binding statement mentioned in the introduction. Precisely, we say that a potential $v$ can (strictly) bind $n$ electrons if the ground-state energy $E_n(v,w)$ belongs to the discrete spectrum of
\[ H_n(v,w) = \sum_{j=1}^n -\Delta_{x_j} + v(x_j) + \sum_{j<k}^n w(|x_j-x_k|). \]
In other words, $v$ binds $n$ electrons if there exists a $n$-particle bound state of $H_n(v,w)$. With this definition, we then have the following corollary.

\begin{corollary}[Binding fewer electrons]\label{cor:binding} For $w$ satisfying the previous assumption and $n\in \N$, any potential $v\in \mH^{-1}_{\rm unif}(\R;\R)$ that can bind $n$ electrons can also bind $n-1$ electrons.
\end{corollary}

\section{Ground-state energy convexity in lattice systems} The goal of this section is to prove that the ground-state energy is convex in the particle number for lattice systems with convex and non-increasing interactions. To state this result precisely, let us introduce some notation.

First, for a fixed number $M\in \N$ of lattice sites, we denote the index set by
\begin{align}
    \mathcal{I}_M \coloneqq \{1,...,M\},
\end{align}
and define the fermionic $n$-particle space as
\[ \mathfrak{h}_0 \coloneqq \C, \qquad \mathfrak{h}_n \coloneqq \wedge^n \ell^2(\mathcal{I}_M) , \qquad n\in \N, \quad n \leq M.\]
We then define the Fock space as
\[ \mathfrak{F}_M \coloneqq \oplus_{n=0}^M \mathfrak{h}_n, \]
and introduce the annihilation $a_j:\mathfrak{h}_{n+1} \rightarrow \mathfrak{h}_n$ and creation $a_j^\ast:\mathfrak{h}_n \rightarrow \mathfrak{h}_{n+1}$ operators at site $j$ via
\[ (a_j \Psi)(j_1,...,j_n) = \sqrt{n+1} \Psi(j,j_1,j_2,...,j_n) \qquad   \inner{a_j^\ast \Psi, \Phi}_{\mathfrak{h}_{n+1}} = \inner{\Psi, a_j \Phi}_{\mathfrak{h}_n}.  \]
For an interaction potential $w: (0,\infty)\rightarrow \R$, an external potential $v\in \R^M$, and a grid spacing $\delta >0$, we define the Hamiltonian 
\[ H_{n,\delta}(v,w)  \coloneqq \sum_{j,k=1}^M \left((T_\delta)_{i,j} + v_i \delta_{ij}\right) a_i^\ast a_j  + \sum_{i < j}^M w(\delta |i-j|) a_i^\ast a_j^\ast a_j a_i \qquad \text{on $\mathfrak{h}_n$,} \]
where $T_\delta$ is the positive discrete Laplacian
\[ (T_\delta)_{i,j} = \begin{dcases} -\delta^{-2}, &\quad \mbox{if $|i-j| = 1$,} \\
2 \delta^{-2}, \qquad &i=j,\\
0, \quad &\mbox{otherwise.} \end{dcases} \]
We then denote the ground-state energy of $H_{n,\delta}(v,w)$ by
\[ E_{n,\delta} \coloneqq \min \left\{ \inner{\Psi, H_{n,\delta}(v,w) \Psi}_{\mathfrak{h}_n} : \Psi \in \mathfrak{h}_n, \quad \norm{\Psi}_{\mathfrak{h}_n} = 1  \right\}. \]

The goal of this section is then to prove the following result.
\begin{theorem}[Ground-state energy convexity on the lattice] \label{thm:discrete convexity} Let $M\in \N$, $\delta>0$, $v\in \R^M$ and suppose that $w:(0,\infty)\rightarrow [0,\infty)$ is convex and non-increasing. Then we have
\[  E_{n+1,\delta} + E_{n-1,\delta} \geq 2E_{n,\delta},\qquad 1\leq n \leq M-1,\]
with the convention $E_{0,\delta} = 0$.
\end{theorem}

\subsection{Configurations and height functions} For the proof of Theorem~\ref{thm:discrete convexity}, it will be convenient to work with the representation of canonical lattice determinants in terms of height functions. Hence, let us introduce the necessary notation.

First, for any set
\[ S\subset \mathcal{I}_M \qquad \text{with} \qquad |S| = \# S = n, \]
we can define the Slater determinant 
\[ e_S = \wedge_{i\in S} e_i  \in \mathfrak{h}_{n}, \qquad \text{where the orbital functions}\qquad e_i(j) \coloneqq \delta_{i,j} \]
are the canonical basis in $\ell^2(\mathcal{I}_M)$ and the $\wedge$-product is ordered in increasing index order. We shall call any such set $S$ a configuration. Since the set of Slater determinants 
\[ \{e_S : S\subset \{1,..., M\},\quad  |S| = n \} \]
forms an orthonormal basis of $\mathfrak{h}_n$, we can represent any operator $A \in \mathcal{B}(\mathfrak{h}_n)$ by a matrix  
\[ A\in \C^{\binom{M}{n}\times \binom{M}{n}} \quad\text{with entries}\quad A(S,S') = \inner{e_S, A e_{S'}}, \qquad S,S' \subset \mathcal{I}_M, |S| = |S'| = n.   \]

Next, we say that a function $h: \{0,..., M\}  \rightarrow \N_0$ is an admissible height if it satisfies
\[ h_0 = 0, \qquad \text{and}\qquad 0\leq h_{i+1} - h_i\leq 1, \qquad \text{for any $i \in \{0, 1,....,M-1\}$.} \]
We denote the set of all admissible height functions by $\mathcal{A}$. Moreover, for $n\leq M$, we define the set of admissible heights of $n$ particle configurations as
\begin{align}
    \mathcal{A}_n \coloneqq \{ h \in \mathcal{A} : h_M = n \}. \label{eq:n-particle weights.} 
\end{align}
The next proposition then summarizes the connection between the configuration and height function representations. 
\begin{proposition}[From configurations to height functions] \label{prop:height functions} The following holds:
\begin{enumerate}[label=(\roman*)]
\item (Height representation) The map defined as
\[ S\subset \{1,..., M\} \rightarrow h_S(i) = \# \{ j \in S : j \leq i \} \]
is a bijection from the power set $\mathcal{P}(\mathcal{I}_M) = \{S: S\subset \mathcal{I}_M \}$ to the set of admissible heights $\mathcal{A}$.
\item \label{it:averaged configurations} For any $S, S' \in \mathcal{P}(\mathcal{I}_M)$ with $|S'| = |S| + 2$, the lower and upper average heights 
\[ h_+ \coloneqq \left\lceil \frac{h_S + h_{S'}}{2} \right\rceil,\qquad h_- \coloneqq \left \lfloor \frac{h_S + h_{S'}}{2} \right\rfloor, \]
where the ceiling and floor functions are understood coordinatewise, are admissible height functions of configurations with $|S|+1$ particles. We refer to the configurations $S_+$ and $S_-$ such that $h_+ = h_{S_+}$ and $h_- = h_{S_-}$ as the upper and lower averaged configurations.
\end{enumerate}
\end{proposition}

\begin{proof} The proof is immediate from the definitions.
\end{proof}

\subsection{Proof of lattice ground-state convexity}

We now need two lemmas to prove Theorem~\ref{thm:discrete convexity}. The first lemma is due to Di Marino, Lewin, and Nenna \cite{DiMarinoLewinNenna2025}, and shows that the sum of the interaction energies of the averaged configurations is not larger than the sum of the interaction energies of the initial configurations. 
\begin{lemma}[Interaction average inequality] \label{lem:interaction estimate}
Let $w:\R_+ \rightarrow \R_+$ be convex and non-increasing. Then for any $S,S' \subset \mathcal{I}_M$ with $|S'| = |S|+2$ we have
\begin{align} \sum_{i<j \in S_-} w(\delta |i-j|) + \sum_{i<j \in S_+} w(\delta |i-j|) \leq \sum_{i< j \in S} w(\delta |i-j|) + \sum_{i<j \in S'} w(\delta|i-j|), \label{eq:interaction inequality} \end{align}
where $S_+$ and $S_-$ are the averaged configurations defined in Proposition~\ref{prop:height functions}~\ref{it:averaged configurations}. 
\end{lemma}
\begin{proof} Let us briefly explain how to pass from \cite[Lemma 5.2]{DiMarinoLewinNenna2025} to the statement here. Let $j_1\leq j_2 ... \leq j_{2n+2}$ be the elements of $S$ and $S'$ ordered in non-decreasing order and such that each element in $S\cap S'$ appears twice. Then, by the definition of $h_+$ and $h_-$ we have
\[ h_+(j_k) - h_+(j_k-1) + h_-(j_k) - h_-(j_k-1) = h_S(j_k) - h_S(j_k-1) + h_{S'}(j_k) - h_{S'}(j_k-1) \]
and
\[ h_- \leq h_+ \leq h_- +1. \]
From the first equation, it follows that any $j_k$ belongs to either $S_+$ or $S_-$, and that $S_+ \cap S_- = S \cap S'$. On the other hand, the second equation implies that we cannot have two consecutive elements $j_k$ and $j_{k+1}$ simultaneously in $S_+$ nor in $S_-$ (unless one of them appears twice, in which case it belongs to both sets). From these observations, we conclude that
\[ S_+ = \{j_1,j_3,j_5,..., j_{2n+1}\} \quad\text{and}\quad S_- = \{j_2,j_4,...,j_{2n+2}\} \]
which are the odd and even partitions appearing in \cite[Lemma 5.2]{DiMarinoLewinNenna2025}. Therefore, their lemma implies~\eqref{eq:interaction inequality}. 
\end{proof}

The second lemma allows us to pass from a pointwise bound on the products of given functions to bounds on the products of their sums. This inequality can be seen as a variant of the Pr\'ekopa-Leindler inequality and was proved in \cite[Theorem 1.4]{KlartagLehec2019}. For simplified proofs we refer to \cite[Theorem 3]{GozlanRobertoSamsonTetali2021} and \cite[Theorem 1.2]{HalikiasKlartagSlomka2021}. 
\begin{lemma}\label{lem:midpoint inequality}
Let $F,G,H,I: \Z^d \rightarrow [0,\infty)$ be functions with finite support in $\Z^d$ and suppose that
\[ F(x) G(y) \leq H\left(\left\lceil \frac{x+y}{2}\right\rceil\right) I\left(\left\lfloor \frac{x+y}{2} \right \rfloor \right), \quad x,y \in \Z^d. \]
Then we have
\begin{align}
    \label{eq:midpoint inequality} \left(\sum_{x\in \Z^d} F(x)\right) \left(\sum_{y \in \Z^d} G(y)\right) \leq \left( \sum_{x\in \Z^d} H(x) \right) \left(\sum_{y\in \Z^d} I(y) \right) . 
\end{align}
\end{lemma}

With these estimates at hand, we can now proceed with the proof of Theorem~\ref{thm:discrete convexity}.

\begin{proof}[Proof of Theorem~\ref{thm:discrete convexity}]
By Proposition~\ref{prop:height functions}, we can represent the Hamiltonian $H_{n,\delta}(v.w)$ in the basis of admissible height functions $h\in \mathcal{A}_n$. In this basis we have
\[ H_{n,\delta}(h,h') = D_n(h) \delta_{h,h'} + F_n(h,h'),\]
where the diagonal part is 
\[ D_n(h) = 2\delta^{-2} n + \sum_{i\in S_h} v_i  + \sum_{i<j \in S_h} w(\delta |i-j|) \]
and the off-diagonal part is
\[ F_n(h,h') = \begin{dcases} - \delta^{-2} \quad &\mbox{if $|h-h'|_1 = 1$,}\\
0, \quad &\mbox{otherwise.} \end{dcases} \]
where $|h-h'|_1 \coloneqq \sum_{j=1}^{M-1} |h_j - h_j'|$ is the $\ell^1$ norm. The formula for $F_n$ follows because $S,S' \subset \mathcal{I}_M$ with $|S| = |S'| = n$ are nearest neighbors if and only if $|h_S - h_{S'}|_1 = 1$. Then note that, since
\[ \sum_{i \in S} v_i + \sum_{j\in S'} v_j = \sum_{i\in S_+} v_i + \sum_{j\in S_-} v_j, \qquad S,S' \subset \mathcal{I}_M,\]
by Lemma~\ref{lem:interaction estimate} we have
\begin{align}
    D_{n-1}(h) + D_{n+1}(h') \geq D_n\left(\left\lfloor \frac{h+h'}{2}\right\rfloor\right) + D_n\left(\left\lceil \frac{h+h'}{2}\right\rceil \right) \quad \text{for $h\in \mathcal{A}_{n-1}, h' \in \mathcal{A}_{n+1}$.} \label{eq:diagonal est} 
\end{align}

We now define the matrix $K_{n,t}$ with elements
\[ K_{n,t}(h,h') \coloneqq e^{-t D_n(h)} (t\delta^{-2})^{|h-h'|_1}. \]
Then, since $\ee^{-t x} = 1 - t x + \mathcal{O}(t^2)$ for any $x\in \R$, we have
\[ K_{n,t} = I - t H_{n,\delta}  + \mathcal{O}(t^2). \]
Consequently
\begin{align} \lim_{\ell \rightarrow \infty} (K_{n,\beta/\ell})^\ell  = \mathrm{e}^{-\beta H_{n,\delta} } ,\quad \text{so in particular} \quad 
\lim_{\ell \rightarrow \infty} \mathrm{Tr} \,K_{n,\beta/\ell}^\ell = \mathrm{Tr}_{\mathfrak{h}_n} \, \mathrm{e}^{-\beta H_{n,\delta}}. 
 \label{eq:partition function convergence} \end{align}
Therefore, it suffices to prove that 
\begin{align}
  (\mathrm{Tr} \, K_{n-1,\beta/\ell}^\ell)  (\mathrm{Tr} \, K_{n+1,\beta/\ell}^\ell) \leq (\mathrm{Tr} \, K_{n,\beta/\ell}^\ell)^2 \qquad\text{for $\beta>0$ and $\ell \in \N$ large enough.}  \label{eq:goal est}
\end{align}
Indeed, if this inequality holds, we can use~\eqref{eq:partition function convergence} to obtain
\[ (\mathrm{Tr}_{\mathfrak{h}_{n-1}} \, \mathrm{e}^{-\beta H_{n-1,\delta}}) (\mathrm{Tr}_{\mathfrak{h}_{n+1}} \, \mathrm{e}^{-\beta H_{n+1,\delta}}) \leq (\mathrm{Tr}_{\mathfrak{h}_n} \, \mathrm{e}^{-\beta H_{n,\delta}})^2. \]
Hence, taking the logarithm and using that
\[ -\lim_{\beta \rightarrow \infty} \frac{1}{\beta} \log \mathrm{Tr}_{\mathfrak{h}_n} (e^{-\beta H_{n,\delta}}) = E_{n,\delta}
\]
completes the proof.

To prove~\eqref{eq:goal est}, we first note that for $0\leq t \leq \delta^2$,
\begin{align} K_{n-1,t}(h,g) K_{n+1,t}(h',g') \leq K_{n,t}(h_+, g_+) K_{n,t}(h_-,g_-), \quad h,g \in \mathcal{A}_{n-1}, h',g' \in \mathcal{A}_{n+1},\label{eq:multiplication inequality} \end{align}
where
\[ h_+ \coloneqq \left\lceil \frac{h+h'}{2}\right\rceil,\quad  h_-\coloneqq \left\lfloor \frac{h+h'}{2}\right \rfloor, \quad g_+ \coloneqq \left\lceil \frac{g+g'}{2}\right\rceil,\quad  g_-\coloneqq \left\lfloor \frac{g+g'}{2}\right \rfloor. \]
Indeed, on one hand, by~\eqref{eq:diagonal est} we have
\begin{align} \ee^{-t \left(D_{n-1}(h) + D_{n+1}(h')\right)} \leq \ee^{-t \left(D_n(h_+) + D_n(h_-)\right)}. \label{eq:diagonal est2} \end{align}
On the other hand, note that, setting
\[ b \coloneqq h_j - g_j ,\quad b'\coloneqq h_j' -g_j',\quad b_- = \left\lfloor \frac{h_j+h_j'}{2} \right\rfloor - \left \lfloor \frac{g_j + g_j'}{2}\right \rfloor, \quad b_+ \coloneqq \left\lceil \frac{h_j+h_j'}{2} \right \rceil - \left \lceil \frac{g_j+g_j'}{2}\right\rceil, \]
we have the set identity
\[ \{b_+, b_-\} = \left\{ \left\lfloor \frac{b+b'}{2} \right\rfloor, \left \lceil \frac{b+b'}{2}\right \rceil \right\}. \]
This holds because $b_+$ and $b_-$ are integers and satisfy $b_- + b_+ = b+b'$ and $|b_+ - b_-|\leq 1$. From this set identity we obtain
\[ |b_+| + |b_-| \leq |b| + |b'|. \]
In particular,
\[ |h_+ - g_+|_1 + |h_- - g_-|_1 \leq |h-g|_1 + |h'-g'|_1 , \]
and therefore
\[ (t\delta^{-2})^{|h-g|_1} (t \delta^{-2})^{|h'-g'|_1} \leq (t\delta^{-2})^{|h_+-g_+|_1} (t\delta^{-2})^{|h_--g_-|_1},\quad  \text{whenever $t\delta^{-2} \leq 1$.} \]
Combining this inequality with \eqref{eq:diagonal est2} then yields~\eqref{eq:multiplication inequality}.

To conclude the proof, we now apply Lemma~\ref{lem:midpoint inequality} to the functions $G_n, G_{n-1}, G_{n+1}: \Z^{M\times \ell} \rightarrow [0,\infty)$ defined as
\begin{align*} &G_r(\{h_i^{(j)}\}_{j \leq  \ell, i \leq M}) \coloneqq \begin{cases}  \prod_{j=1}^\ell K_{r,\beta/\ell}(h^{(j)}, h^{(j+1)}) , &\text{if $h^{(j)} \in \mathcal{A}_r$ for any $j$,} \\
0, &\text{otherwise,} \end{cases}  \end{align*}
with the convention $h^{(\ell+1)} = h^{(1)}$. Precisely, note that 
\[ G_{n-1}(x) G_{n+1}(y) \leq G_n\left(\left\lfloor \frac{x+y}{2}\right\rfloor\right) G_n\left(\left\lceil \frac{x+y}{2}\right \rceil\right), \quad \text{for $x,y \in \Z^{M \times \ell}$ by~\eqref{eq:multiplication inequality},} \]
and the traces in~\eqref{eq:goal est} can be written as
\begin{align*} \mathrm{Tr}\, K_{n,\beta/\ell}^\ell &= \sum_{(h^{(1)},...., h^{(\ell)}) \in (\mathcal{A}_n)^\ell}\prod_{j=1}^\ell K_{n,\beta/\ell}(h^{(j)}, h^{(j+1)}) =  \sum_{(h^{(1)},...., h^{(\ell)}) \in (\mathcal{A}_n)^\ell} G_n(\{h_i^{(j)}\}) \\
&=\sum_{x\in \Z^{M\times \ell}} G_n(x).  \end{align*}
Therefore, applying~\eqref{eq:midpoint inequality} to $F = G_{n-1}, G = G_{n+1}, H = I= G_n$ yields~\eqref{eq:goal est}.
\end{proof}

\section{From the discrete to the continuum limit} 

Our goal in this section is to prove Theorem~\ref{thm:GS convexity}. 

\subsection{Discrete to continuum limit with regular potentials on bounded intervals} We start by passing the results of Theorem~\ref{thm:discrete convexity} to the continuum limit on a bounded interval and assuming continuous and bounded external and interaction potentials. 
\begin{theorem}[Ground-state convexity for regular potentials on the interval] \label{thm:dirichlet GS convexity} Let $I = (a,b) \subset \R$ be a bounded interval, $v\in C(\overline{I};\R)$, and $w\in C([0,b-a];[0,\infty))$ be convex and non-increasing function. Then the ground-state energy
\[ E_n(v,w)\coloneqq \min_{\Psi \in \mH^1_0(I^n) \cap \wedge^n \mL^2(I)\setminus \{0\}} \left\{ \frac{\int_{I^n} |\nabla \psi|^2(X) + \left(\sum_{j=1}^n v(x_j) + \sum_{j<k} w(|x_j-x_k|)\right) |\psi(X)|^2 \mathrm{d} X}{\norm{\Psi}_{\mL^2}^2} \right\} \]
satisfies
\[ E_{n+1}(v,w) + E_{n-1}(v,w) \geq 2 E_n(v,w) ,\qquad \mbox{for any $n\in \N$.}\]
\end{theorem}

For the proof of this result, let us recall the following well-known interpolation estimates.

\begin{lemma}[Finite element interpolation estimates] \label{lem:interpolation estimates} Let $I=(a,b)$ be a bounded interval, $M\in \N$ and set
\[\delta  \coloneqq \frac{b-a}{M+1} \quad \text{and}\quad y_j \coloneqq a + j \delta. \]
For any configuration $S \subset \{1,..., M\}$ such that $|S| = n$, let $\{j_1, ..., j_n\} = S$ denote the elements listed in increasing order $j_1 < j_2 ... < j_n$ and define the interpolation functions
\[ \eta_{S,\delta}(x_1,...,x_n) = \sum_{\sigma \in \mathcal{S}_n} \mathrm{sign}(\sigma) \prod_{r=1}^n \phi\left(\frac{x_{\sigma(r)} -y_{j_r}}{\delta}\right), \qquad \phi(x) \coloneqq \max\{0, 1- |x|\}. \]
Let $J_\delta: \C^{\binom{M}{n}} \rightarrow \mH^1_0(I^n) \cap \wedge^n \mL^2(I)$ be the interpolation operator
\begin{align} (J_\delta u)(x) = \sum_{|S| = n} u(S) \eta_{S,\delta}(x).\label{eq:interpolation operator} \end{align}
Then, the following estimates hold:
\begin{align}
    &0\leq  \delta^n n!  \sum_{S} |u(S)|^2- \norm{J_\delta u}_{L^2(I^n)}^2  \leq \frac{\delta^{n+2} n!}{6} \inner{u, \hat{T}_\delta u}_{\C^{\binom{M}{n}}}, \label{eq:L2 norm bound} \\
    &\norm{\nabla J_\delta u}_{\mL^2(I^n)}^2 \leq \delta^n n! \inner{u, \hat{T}_\delta u}_{\C^{\binom{M}{n}}}. \label{eq:kinetic energy bound}
\end{align}
Moreover, for any symmetric $0\leq V\in C(\overline{I^n})$, we have
\begin{align}
    &\left|\inner{J_\delta u, V J_\delta u} - n! \delta^n \sum_{|S|=n}  V(S) |u(S)|^2 \right| \lesssim \mu_V(\delta) \delta^n n! \sum_{|S| = n} |u(S)|^2 + \norm{V}_\infty n! \delta^{n+2} \inner{u, \hat{T}_\delta u},  \label{eq:potential energy bound}
\end{align}
where $\hat{T}_\delta = \sum_{i,j} (T_{\delta})_{i,j} a_i^\ast a_j$ is the $n$-particle lift of the discrete Laplacian $T_\delta$, $V(S) = V(y_{j_1},...,y_{j_n})$ for $S = \{j_1,...,j_n\}$ (where the order does not matter as $V$ is symmetric) and
\[ \mu_V(\delta) \coloneqq \norm{\sqrt{V}}_\infty \sup \left\{|\sqrt{V}(x) - \sqrt{V}(y)| : x,y\in I^n , \quad |x_j-y_j| \leq \delta, \quad\text{for all $j\leq n$} \right\}\]
\end{lemma}

\begin{proof} Estimates of this kind are fairly standard in the literature on finite difference method, see, e.g., \cite[Chapter 6]{Weinberger1974}. For the sake of completeness, and as we could not find exactly the statements we need elsewhere, we present the proof here.

    First, straightforward calculations show that 
    \begin{align*} &A_{i,j} \coloneqq \int_\R \phi((x-y_i)/\delta) \phi((x-y_j)/\delta) \mathrm{d} x = \begin{cases} \frac{2\delta}{3} , \quad &\mbox{if $j=i$,} \\
    \frac{\delta}{6}, \quad &\mbox{if $|j-i| = 1$,} \\
    0, \quad &\mbox{otherwise,} \end{cases} \\
    &B_{i,j} \coloneqq \int_\R \partial_x(\phi((x-y_i)/\delta)) \partial_x(\phi((x-y_j)/\delta)) \mathrm{d} x = \begin{cases} \frac{2}{\delta}, \quad &\mbox{for $i=j$,} \\ -\frac{1}{\delta}, \quad &\mbox{for $|i-j|=1$,} \\ 0, \quad &\mbox{otherwise.} \end{cases} \end{align*}
    Therefore, we have the identities
    \[ A = \delta\left(I - \frac{\delta^2}{6} T_\delta \right)\quad\text{and}\quad B = \delta T_\delta, \qquad \text{where $T_\delta$ is the positive discrete Laplacian.} \]
    Now for $u \in \C^{\binom{M}{n}}$, we define the extended vector $\tilde{u} \in \C^{M^n}$ as
    \[ \tilde{u}(j_1,...,j_n) = \begin{cases} \det \sigma_J u(\{j_1,...,j_n\}), \quad &\mbox{if $\# \{j_1,...,j_n\} = n$,} \\ 0, \quad &\mbox{otherwise,} \end{cases} \]
    where $\sigma_J$ is the unique permutation matrix such that $\sigma_J(j_1,...,j_n)$ is the non-decreasing rearrangement of $(j_1,...,j_n)$. Note that
    \begin{align} \sum_{J\in \mathcal{I}_M^n} |\tilde{u}(J)|^2 = n! \sum_{|S| = n} |u(s)|^2. \label{eq:permutation formula} \end{align}
    Moreover, using this definition of $\tilde{u}$ we can write 
    \[ (J_\delta u)(x_1,x_2,...,x_n) = \sum_{J \in \{1,..,M\}^n} \tilde{u}(J) \prod_{\ell=1}^n \phi\left(\frac{x_\ell - y_{j_\ell}}{\delta}\right). \]
    Consequently
    \begin{align*} \norm{J_\delta u}_{\mL^2(I^n)}^2 &= \sum_{J,K \in \{1,...,M\}^n} \tilde{u}(J) \overline{\tilde{u}(K)} \prod_{\ell=1}^n A_{j_\ell, k_\ell} = \inner{\tilde{u}, \otimes^n A \tilde{u}}_{\C^{M^n}} = \delta^n \inner{\tilde{u}, \otimes^n \left(1-\frac{\delta^2}{6} T_\delta\right) \tilde{u}}_{\C^{M^n}}.  \end{align*}
    Therefore, using that $0\leq 1- \frac{\delta^2}{6} T_\delta \leq 1$ we see that $1- \frac{\delta^2}{6}\hat{T_\delta} \leq \otimes^n (1-\frac{\delta^2}{6} T_\delta) \leq 1$, which together with the above formula and~\eqref{eq:permutation formula} yields~\eqref{eq:L2 norm bound}. 

    For the kinetic energy bound~\eqref{eq:kinetic energy bound}, note that
    \begin{align*} \norm{\partial_{x_1} J_\delta u}_{\mL^2(I^n)}^2 &= \sum_{J,K\in \{1,..., M\}^n } \tilde{u}(J) \overline{\tilde{u}(K)} B_{j_1,k_1} \prod_{\ell=2}^n A_{j_\ell, k_\ell} = \inner{\tilde{u}, B \otimes^{n-1} A, \tilde{u}}_{\C^{M^n}} \\
    &= \delta^n \inner{\tilde{u}, T_\delta \otimes^{n-1} (I-\frac{\delta^2}{6} T_\delta), \tilde{u}}_{\C^{M^n}} . \end{align*}
    Hence, summing this identity for all coordinates and using again $I-\delta^2 T_\delta/6 \leq I$ and~\eqref{eq:permutation formula} we obtain~\eqref{eq:kinetic energy bound}. 

    For the potential estimate, first note that, since $V$ is symmetric and $\phi$ non-negative, we find
    \begin{multline*} \left|\int_{I^n} \left(V(x) - \sqrt{V(y_{j_1},...,y_{j_\ell})}\sqrt{V(y_{k_1},...,y_{k_n})} \right) \prod_{\ell=1}^n \phi\left(\frac{x_\ell - y_{j_\ell}}{\delta}\right) \phi\left(\frac{x_\ell - y_{k_\ell}}{\delta}\right) \mathrm{d} X \right| \\
    \leq 2\norm{\sqrt{V}}_\infty \sup_{|x-y|_\infty \leq \delta} |\sqrt{V}(x) - \sqrt{V}(y)| \prod_{\ell=1}^n A_{j_\ell,k_\ell} =: 2\mu_V(\delta) \prod_{\ell=1}^n A_{j_\ell,k_\ell}.   \end{multline*}
    Using this estimate and~\eqref{eq:L2 norm bound}, we obtain
    \begin{align} \left|\inner{J_\delta u, V  J_\delta u} - 
\norm{J_\delta (\sqrt{V} u)}_{\mL^2}^2 \right| \leq 2\mu_V(\delta) \inner{|\tilde{u}|, \otimes^n A |\tilde{u}|} \leq 2\mu_V(\delta) n! \delta^n \sum_{|S|=n} |u(S)|^2, \label{eq:some first est} \end{align}
    where we slightly abused the notation by denoting $\sqrt{V} \in \C^{M^n}$ as $\sqrt{V}(J) = V(y_{j_1},..., y_{j_n})$. Moreover, by~\eqref{eq:L2 norm bound} and~\eqref{eq:permutation formula} we have
    \begin{align} 0\leq \delta^n \inner{\sqrt{V} \tilde{u}, \sqrt{V} \tilde{u}}_{\C^{M^n}} - \norm{J_\delta (\sqrt{V} u)}_{\mL^2}^2  \leq  \frac{\delta^{n+2}}{6} \inner{\sqrt{V} \tilde{u}, \hat{T}_\delta \sqrt{V} \tilde{u}}. \label{eq:some intermediate est} \end{align}
    Furthermore, using the identity 
    \begin{multline*} (\hat{T_\delta} \sqrt{V} u)(J) = \sqrt{V}(J) (\hat{T}_\delta \tilde{u})(J) \\
    + \delta^{-2} \sum_{k=1}^n \tilde{u}(J+e_k) \left(\sqrt{V}(J+e_k)-\sqrt{V}(J)\right) + \tilde{u}(J-e_k)\left(\sqrt{V}(J-e_k) - \sqrt{V}(J)\right), \end{multline*}
    we can bound the left-hand side of~\eqref{eq:some intermediate est} by
    \begin{align} \frac{\delta^{n+2}}{6} \inner{\sqrt{V} \tilde{u}, \hat{T}_\delta \sqrt{V} \tilde{u}} \leq n\frac{\delta^{n+2}}{6} \norm{V}_\infty \inner{\tilde{u}, \hat{T}_\delta \tilde{u}} +  \frac{2n \mu_V(\delta)}{6} n! \delta^n  \sum_{|S| = n} |u(S)|^2. \label{eq:last est} \end{align}
    Finally, combining~\eqref{eq:some first est},~\eqref{eq:some intermediate est}, and~\eqref{eq:last est} yields~\eqref{eq:potential energy bound}. 
\end{proof}

We can now proceed with the proof of Theorem~\ref{thm:dirichlet GS convexity}. 
\begin{proof}[Proof of Theorem~\ref{thm:dirichlet GS convexity}] Set
    \[ \delta \coloneqq \frac{b-a}{M+1}, \qquad y_j \coloneqq  a+ j \delta ,  \]
   and define the lattice Hamiltonian 
    \[ H_{n,\delta} \coloneqq \sum_{i,j=1}^M \left((T_\delta)_{i,j} + v(y_i) \delta_{i,j} \right) a_i^\ast a_j + \sum_{i< j} w(\delta|i-j|) a_i^\ast a_j^\ast a_j a_i\quad \text{acting on $\mathfrak{h}_n$.} \]
    By Theorem~\ref{thm:discrete convexity}, we already have the convexity of the ground-state energy $E_{n,\delta}(v,w)$ of the discrete system. Therefore, it suffices to show that
    \[ \lim_{\delta \downarrow 0} E_{n,\delta}(v,w) = E_n(v,w). \]

    For the lower bound, we let $\psi \in C^\infty_c(I^n) \cap \wedge^n L^2(I)$ and define, for any configuration $S\subset \{1,..., M\}$ with $|S|=n$, the vector
    \[ \psi_\delta(S) \coloneqq \psi(x_{j_1},x_{j_2},...,x_{j_n}), \qquad \text{where $j_1< j_2 ... < j_n$ and $S = \{j_1,..., j_n\}$.} \]
    Then, since $v$ and $w$ are continuous and bounded and $\psi$ has zero trace on the boundary $\partial I^n$, we have convergence of the Riemann sums 
    \begin{align*} &\delta^n n! \sum_{|S| = n}  H_{n,\delta}(S,S') \overline{\psi_\delta(S')} \psi_\delta(S)\rightarrow \inner{\Psi, (-\Delta + V_n) \Psi}_{\mL^2(I^n)}, \\
    &\delta^n n! \sum_{S\subset\{ 1,..., M\}} |\Psi_\delta(S)|^2\rightarrow \norm{\Psi}_{\mL^2(I^n)}^2, \end{align*}
    as $\delta \downarrow 0$, where we used the shorthand notation 
    \begin{align} V_n(X) \coloneqq \sum_{j} v(x_j) + \sum_{j \neq k} w(|x_j-x_k|). \label{eq:shorthand V_n} \end{align} 
    Thus, since $C^\infty_c(I^n) \cap \wedge^n \mL^2(I)$ is dense in $\mH^1_0(I^n) \cap \wedge^n \mL^2(I)$, we obtain the upper bound
    \[ \limsup_{\delta \downarrow 0} E_{n,\delta}(v,w) \leq E_n(v,w).  \]

    For the opposite inequality, let $u_\delta$ be the ground-state of $H_{n,\delta}$ in the $n$-particle sector $\mathfrak{h}_n$ and assume $u_\delta$ normalized as
    \begin{align}  n! \delta^n \sum_{|S| = n} |u_\delta(S)|^2. \label{eq:normalization u} \end{align}
    Define the wave-functions $J_\delta u_\delta \in \mH^1_0(I^n) \cap \wedge^n L^2(I)$, where $J_\delta$ is the interpolation operator in~\eqref{eq:interpolation operator}. Then, if $\limsup_{\delta} E_{n,\delta} = \infty$, there is nothing to prove. Assuming otherwise, estimate~\eqref{eq:kinetic energy bound} and the normalization of $u_\delta$ implies that
    \begin{align} \limsup \norm{\nabla J_\delta u_\delta}^2 \leq \limsup  n! \delta^n \inner{u_\delta, \hat{T}_\delta u_\delta} \leq \limsup E_{n,h} + \norm{V_n}_\infty < \infty.\label{eq:uniform upper bound} \end{align}
    Thus $J_\delta u_\delta$ is uniformly bounded in $\mH^1$. As $I^n$ is bounded, we can therefore extract a subsequence such that $J_\delta u_\delta \rightarrow \psi$ strongly in $\mL^2$ and weakly in $\mH^1_0$. Estimate~\eqref{eq:L2 norm bound} then implies that $\norm{\psi} = 1$. 
    
    Next, we note that, after possibly shifting $V_n$ by a constant, we can assume $V_n\geq 0$. Moreover, since $V_n$ is bounded and continuous in $\overline{I^n}$, it is uniformly continuous, and therefore
    \begin{align} \lim_{\delta \downarrow 0} \omega_{V_n}(\delta) = 0. \label{eq:V continuity} \end{align} 
    Hence, using estimate~\eqref{eq:potential energy bound}, the uniform bound on the discrete kinetic energy in~\eqref{eq:uniform upper bound} and~\eqref{eq:normalization u}, we get
    \[ \lim_{\delta \downarrow 0} \int V_n |J_\delta u_\delta| - n! \delta^2 \sum_{S} V_n(S) |u_\delta(S)|^2 = 0.\]
    Thus, using this estimate, the weak lower semi-continuity of the $\mH^1$-norm, and~\eqref{eq:kinetic energy bound} we obtain
    \[ E_n(v,w) \leq \norm{\nabla \psi}^2 + \inner{\psi, V_n \psi} \leq \liminf_{\delta\downarrow 0} \norm{\nabla J_\delta u} + \inner{J_\delta u_\delta, V_n J_\delta u_\delta} \leq \liminf E_{n,h}(v,w), \]
    which completes the proof.
\end{proof}

\subsection{Extending to singular potentials on the line}
To complete the proof of Theorem~\ref{thm:GS convexity} we now extend Theorem~\ref{thm:dirichlet GS convexity} to singular potentials and pass to the whole line limit. For this, we use the following simple approximation estimates.

\begin{lemma}[Singular external and interaction potentials] \label{lem:potentials} Let $v\in \mH^{-1}_{\rm unif}(\R)$ and $w:(0,\infty) \rightarrow \R_+$ be convex and non-increasing. Then the following holds:
\begin{enumerate}[label=(\roman*)]
\item For any $\Psi \in \mH^1(\R^n)$ we  have
\begin{align} |\inner{v,\rho_\Psi}| \lesssim \norm{v}_{\mH^{-1}_{\rm unif}} \norm{\Psi}_{\mH^1}^{\frac32} \norm{\Psi}_{\mL^2}^{\frac12}. \label{eq:v form bound} \end{align}
In particular, the form $\Psi \mapsto \inner{v,\rho_\Psi}$ is infinitesimally Laplace form-bounded.
\item For any bounded interval $I=(a,b) \subset \R$, there exists $v_{\ell} \in C^\infty_c(\R)$ such that
\begin{align} \lim_{\ell \rightarrow \infty} |||v-v_\ell||| =0 ,\quad \text{where}\quad |||v-v_\ell||| \coloneqq \sup_{\substack{\Psi \in \mH^1_0(I^n) \\ \norm{\Psi}_{\mH^1(I^n)} \leq 1}} \left| \int_I v_\ell(x) \rho_\Psi(x) \mathrm{d} x - \inner{v, \rho_\Psi}\right| = 0. \label{eq:potential approximation} \end{align}
\item \label{it:interaction approximation} The interaction potential $w_\varepsilon(t) \coloneqq w(t+\varepsilon)$ is convex, non-increasing and satisfies 
\[ w_\varepsilon(x) \uparrow w(x) \qquad \text{as $\varepsilon \downarrow 0$.} \]
\end{enumerate}
\end{lemma}

\begin{proof}
    For estimate~\eqref{eq:v form bound}, note that from Cauchy-Schwarz, 
    \[ \norm{\nabla \sqrt{\rho_\Psi}}_{\mL^2(I)}^2 \leq n \norm{\nabla \Psi}_{\mL^2(I\times \R^{n-1})}^2,\quad \mbox{for any interval $I\subset \R$.} \]
    Hence, by Cauchy-Schwarz and the 1D GNS inequality, for any unit interval $I\subset \R$, we have
    \begin{align*} \norm{\nabla \rho}_{\mL^2(I)}^2 &\leq \norm{\nabla \sqrt{\rho_\Psi}}_{\mL^2(I)}^2 \norm{\sqrt{\rho_\Psi}}_{\mL^\infty(I)}^2 \lesssim n\norm{\nabla \Psi}_{\mL^2(I\times \R^{n-1})}^2 \norm{\sqrt{\rho}}_{\mH^1(I)} \norm{\sqrt{\rho}}_{\mL^2(I)} \\
    &\lesssim n^2 \norm{\nabla \Psi}_{\mL^2(I\times \R^{n-1})}^2 \norm{\Psi}_{\mH^1(I\times \R^{n-1})} \norm{\Psi}_{\mL^2(I\times \R^{n-1})}.\end{align*}
    Thus, using the definition of $\mH^{-1}_{\rm unif}(\R)$, a partition of the unity argument, the above estimate, and Cauchy-Schwarz, we find
    \[ |\inner{v,\rho_\Psi} \lesssim \sum_{t\in \Z} \norm{v}_{\mH^{-1}_{\rm unif}} \norm{\Psi}_{\mH^1(I_t\times \R^{n-1})}^{\frac32} \norm{\Psi}_{\mL^2(I_t\times\R^{n-1})}^{\frac12} \leq \norm{v}_{\mH^{-1}_{\rm unif}} \norm{\Psi}_{\mH^1}^{\frac32} \norm{\Psi}_{\mL^2}^{\frac12}. \]

    The approximation estimate~\eqref{eq:potential approximation} follows from estimate~\eqref{eq:v form bound} and the fact that $C^\infty_c(I)$ is dense in $\mH^{-1}(I)$, see, e.g. \cite[Lemma 3.10]{Cor26b}.

    Item~\ref{it:interaction approximation} follows from the fact that any convex function is (locally Lipschitz) continuous on the interior of its domain. 
\end{proof}

\begin{proof}[Proof of Theorem~\ref{thm:GS convexity}]
By Theorem~\ref{thm:dirichlet GS convexity}, the convexity estimate holds in a bounded interval and for bounded external and interaction potentials. In particular, it holds for the ground-state energies $E_n(v_\ell,w_\varepsilon, L)$ of the Hamiltonians
\[ H_n(v_\ell,w_\varepsilon,L) = -\Delta + \sum_{j=1}^n v_\ell(x_j) + \sum_{k\neq j}^n w_\varepsilon(|x_j-x_k|) \quad \text{with form domain} \quad \mH^1_0(I_L^n) \cap \wedge^n \mL^2(I_L), \]
where $I_L \coloneqq (-L, L)$. Therefore, it suffices to show that
\[ \lim_{L\rightarrow \infty} \lim_{\varepsilon \downarrow 0} \lim_{\ell \rightarrow \infty} E_n(v_\ell,w_\varepsilon,L) = E_n(v,w). \]

For the first limit, since $w_\varepsilon\geq 0$, by the Laplace form-bound in~\eqref{eq:v form bound} and Young's inequality,
\begin{align} \inner{\Psi, H_n(v,w_\varepsilon,L) \Psi} \geq \frac12 \norm{\nabla \Psi}_{\mL^2}^2 - C \norm{\Psi}_{\mL^2}^2, \quad \mbox{for $\Psi \in \mH^1_0(I_L^n) \cap \wedge^n \mL^2(I_L)$.} \label{eq:lower kinetic bound} \end{align}
Hence, for any $\Psi \in \mH^1_0(I_L^n) \cap \wedge^n \mL^2(I_L)$, we have 
\begin{align*} |\inner{\Psi, H_n(v_\ell,w_\varepsilon,L) \Psi} - \inner{\Psi, H_n(v,w_\epsilon,L) \Psi}| &\leq |||v-v_\ell|||\, \norm{\Psi}_{\mH^1}^2  \\
&\leq |||v-v_\ell||| \left(2\inner{\Psi, H_n(v,w_\varepsilon,L) \Psi} + C \norm{\Psi}_{\mL^2}^2 \right).
\end{align*}
Hence, we conclude from~\eqref{eq:potential approximation} that $E_n(v_\ell,w_\varepsilon,L) \rightarrow E(v,w_\varepsilon, L)$ as $\ell\rightarrow \infty$.

To pass to the limit in $w$, we first note that
\[ E_n(v,w_\varepsilon, L) \leq E_n(v,w,L) \quad \text{as} \quad w_\varepsilon \leq w.\]
For the converse, we assume that $\limsup E_n(v,w_\varepsilon,L) <\infty$ as otherwise there is nothing to prove by the above estimate. Then, we take a sequence of normalized ground-states\footnote{Note that a ground-state of $H_n(v,w_\varepsilon,L)$ exists because the form domain $\mH^1_0(I_L^n) \cap \wedge^n \mL^2(I_L)$ is compactly embedded in $\mL^2(I_L^n)$ so that $H_n(v,w_\varepsilon,L)$ has discrete spectrum. See, e.g., \cite[Section 3]{Cor26b} for more details.} $\Psi_\varepsilon$ of $H_n(v,w_\varepsilon,L)$. As $0\leq w_\varepsilon \uparrow w$, estimate~\eqref{eq:kinetic energy bound} holds with $C>0$ uniform in $\varepsilon$. In particular, the sequence $\{\Psi_\varepsilon\}$ is $\mH^1$-bounded. As $I_L$ is bounded we can extract a subsequence such that $\Psi_\varepsilon \rightarrow \Psi$ strongly in $\mL^2$ and weakly in $\mH^1$. In particular, $\rho_{\Psi_\epsilon} \rightarrow \rho_\Psi$ weakly in $\mH^1(I_L)$. Therefore, by weak lower semi-continuity of $\mH^1$-norm we obtain
\begin{align} \norm{\nabla \Psi}_{\mL^2}^2 + \inner{v,\rho_\Psi} \leq \liminf_{\varepsilon \downarrow 0} \norm{\nabla \Psi_\varepsilon}_{\mL^2}^2 + \inner{v,\rho_{\Psi_\varepsilon}}. \label{eq:one-body lower} \end{align}
On the other hand, as $w_\varepsilon \rightarrow w$ pointwise and $|\Psi_\varepsilon|^2 \rightarrow |\Psi|^2$ almost everywhere, up to a subsequence, Fatou's lemma yields
\[ \int_{I_L^n} \sum_{j<k} w(|x_j-x_k|) |\Psi(X)|^2 \mathrm{d} X \leq \liminf_{\varepsilon \downarrow 0} \int_{I_L^n} \sum_{j<k} w_\varepsilon(|x_j-x_k|) |\Psi_\varepsilon(X)|^2\mathrm{d} X. \]
Combining these two estimates then shows that $E_n(v,w,L) \leq \liminf E_n(v,w_\varepsilon,L)$. Therefore, $E_n(v,w_\varepsilon,L) \rightarrow E_n(v,w,L)$ as $\varepsilon \downarrow 0$.

For the last limit we argue as follows. On one hand, 
\[ E_n(v,w,L) \geq E_n(v,w) \]
as the latter involves minimizing the same quadratic form as the former but in a larger space. On the other hand, let $\eta \in C^\infty_c([-1,1];[0,1])$ be such that $\eta = 1$ close to zero and $\eta$ is even and non-increasing in the positive axis. Then, for any $\Psi \in \mH^1(\R^n) \cap \wedge^n \mL^2(\R)$, the sequence 
\[ \Psi_L(X) \coloneqq \prod_{j=1}^n \eta(x_j/L) \Psi(X) \]
converges to $\Psi$ in $\mH^1(\R)$ and satisfies $|\Psi_L|^2 \uparrow |\Psi|^2$ point-wise. Therefore, by $\mH^1$-continuity of the external potential form and monotone convergence for the interaction term we find
\[ \limsup_{L\rightarrow \infty} E_n(v,w,L) \leq \limsup_{L\rightarrow \infty} \frac{\inner{\Psi_L, (-\Delta + V_n) \Psi_L}}{\norm{\Psi_L}^2} = \frac{\inner{\Psi, -\Delta + V_n \Psi}}{\norm{\Psi}^2}, \]
where we used again the shorthand notation $V_n$ for the combined potentials, see~\eqref{eq:shorthand V_n}. 
As this holds for any $\Psi \in \mH^1(\R^n) \cap \wedge^n \mL^2(\R)$, the opposite inequality follows by taking the infimum over $\Psi$. 
\end{proof}

\section*{Acknowledgements}

The author is grateful to Mathieu Lewin for exchanging some ideas on how to tackle the problem studied in this paper.

T.~C.~Corso acknowledges funding by the \emph{Deutsche Forschungsgemeinschaft} (DFG, German Research Foundation) - Project number 572811220. 

\section*{AI disclosure}
The results of this manuscript were obtained with the assistance of Large Language Models (LLM). More precisely, the main arguments in the proof of Theorem~\ref{thm:GS convexity} were suggested by an LLM. The main contribution of the author was to understand, simplify, and rewrite these arguments in a coherent form. The paper is entirely written by the author and contains no LLM output. In particular, the author takes full responsibility for the content of the paper.

\section*{Data availability}
No datasets were generated or analyzed during the current study.

\section*{Competing interests}
The authors have no competing interests to declare that are relevant to the content of this article.

\appendix

\printbibliography

\end{document}